\documentclass[aps,prd,nofootinbib,showpacs,twocolumn,10pt]{revtex4}
\usepackage{txfonts}
\usepackage{graphicx}
\usepackage{dcolumn}
\usepackage{bm}
\usepackage{amssymb}
\usepackage{latexsym}
\usepackage{booktabs}
\usepackage[colorlinks, linkcolor=blue, citecolor=blue, urlcolor=blue]{hyperref}

\newcommand{\be}{\begin{equation}}
\newcommand{\ee}{\end{equation}}
\newcommand{\bq}{\begin{eqnarray}}
\newcommand{\eq}{\end{eqnarray}}

\begin{document}

\title{Cosmic age problem revisited in the holographic dark energy model}

\author{Jinglei Cui and Xin Zhang}
\affiliation{Department of Physics, College of Sciences,
Northeastern University, Shenyang 110004, China}

\begin{abstract}
Because of an old quasar APM $08279+5255$ at $z=3.91$, some dark
energy models face the challenge of the cosmic age problem. It has
been shown by Wei and Zhang [Phys. Rev. D {\bf 76}, 063003 (2007)]
that the holographic dark energy model is also troubled with such a
cosmic age problem. In order to accommodate this old quasar and
solve the age problem, we propose in this Letter to consider the
interacting holographic dark energy in a non-flat universe. We show
that the cosmic age problem can be eliminated when the interaction
and spatial curvature are both involved in the holographic dark
energy model.
\end{abstract}

\maketitle


\section{Introduction}\label{sec1}
The fact that our universe is undergoing accelerated expansion has
been confirmed by lots of astronomical observations such as type Ia
supernovae (SNIa)~\cite{Riess98}, large scale structure
(LSS)~\cite{Tegmark04} and cosmic microwave background (CMB)
anisotropy~\cite{Spergel03}. It is the most accepted idea that this
cosmic acceleration is caused by some kind of negative-pressure
matter known as dark energy whose energy density has been dominative
in the universe. The combined analysis of cosmological observations
indicates that the universe today consists of about 70\% dark
energy, 30\% dust matter (cold dark matter plus baryons), and
negligible radiation. The famous cosmological constant $\lambda$
introduced first by Einstein is the simplest candidate for dark
energy. However, the cosmological constant scenario has to face the
so-called ``fine-tuning problem'' and ``cosmic coincidence
problem''~\cite{dereview}. Many dark energy models have been
proposed, while the nature of dark energy is still obscure. Besides
quintessence~\cite{quintessence}, a wide variety of scalar-field
dark energy models have been studied including
$k$-essence~\cite{kessence}, hessence~\cite{hessence},
phantom~\cite{phantom}, tachyon~\cite{tachyon},
quintom~\cite{quintom}, ghost condensate~\cite{ghost}, etc. In
addition, there are other proposals on dark energy such as
interacting dark energy models~\cite{intde}, brane world
models~\cite{brane}, Chaplygin gas models~\cite{cg}, Yang-Mills
condensate models~\cite{YMC}, and so on.

The dark energy problem is essentially an issue of quantum gravity,
owing to the concern of the vacuum expectation value of some quantum
fields in a universe governed by gravity. However, by far, we have
no a complete theory of quantum gravity yet. So, it seems that we
have to consider the effects of gravity in some effective quantum
field theory in which some fundamental principles of quantum gravity
could be taken into account. It is commonly believed that the
holographic principle \cite{hp} is just a fundamental principle of
quantum gravity. Based on the effective quantum field theory, Cohen
et al. \cite{r16} pointed out that the quantum zero-point energy of
a system with size $L$ should not exceed the mass of a black hole
with the same size, i.e., $L^3\Lambda^4\leq LM_{Pl}^2$, where
$\Lambda$ is the ultraviolet (UV) cutoff of the effective quantum
field theory, which is closely related to the quantum zero-point
energy density, and $M_{Pl}\equiv 1/\sqrt{8\pi G}$ is the reduced
Planck mass. This observation relates the UV cutoff of a system to
its infrared (IR) cutoff. When we take the whole universe into
account, the vacuum energy related to this holographic principle can
be viewed as dark energy (its energy density is denoted as
$\rho_{\Lambda}$ hereafter). The largest IR cutoff $L$ is chosen by
saturating the inequality, so that we get the holographic dark
energy density
\begin{equation}
\label{eq1}\rho_{\Lambda}=3c^2M_{Pl}^2L^{-2}
\end{equation}
where $c$ is a numerical constant characterizing all of the
uncertainties of the theory, and its value can only be determined by
observations. If we take $L$ as the size of the current universe,
say, the Hubble radius $H^{-1}$, then the dark energy density will
be close to the observational result. However, Hsu \cite{r17}
pointed out that this yields a wrong equation of state for dark
energy. Subsequently, Li \cite{Li04} suggested to choose the future
event horizon of the universe as the IR cutoff of this theory. This
choice not only gives a reasonable value for dark energy density,
but also leads to an accelerated universe. Moreover, the cosmic
coincidence problem can also be explained successfully in this
model, provided that the inflation lasts for more than 60 $e$-folds.
Most recently, a calculation of the Casimir energy of the photon
field in a de Sitter space is performed \cite{Li:2009pm}, and it is
a surprising result that the Casimir energy is indeed proportional
to the size of the horizon (the usual Casimir energy in a cavity is
inversely proportional to the size of the cavity), in agreement with
the holographic dark energy model.

Up to now, the holographic dark energy model has been tested by
various observational data including SNIa~\cite{r38},
SNIa+BAO+CMB~\cite{holodata,Li:2009zs}, X-ray gas mass fraction of
galaxy clusters~\cite{r40}, differential ages of passively evolving
galaxies~\cite{r41}, Sandage-Leob test~\cite{r42}, and so on
\cite{Shen:2004ck}. These analyses show that the holographic dark
energy model is consistent with the observational data. However, Wei
and Zhang~\cite{hao} used some old high redshift objects (OHROs) to
test the holographic dark energy model and found that the original
holographic dark energy model can be ruled out unless a lower Hubble
constant (e.g., $h=0.56$) is taken. So, according to
Ref.~\cite{hao}, there is a cosmic age crisis in the holographic
dark energy model.

In fact, many dark energy models are in the face of such a cosmic
age problem. In history, the cosmic age problem has been focused in
cosmology for several times. At present, the cosmic age crisis
coming from some OHROs appears again in cosmological models, even
though dark energy is involved in the models. In cosmology there is
a very basic principle that the universe cannot be younger than its
constituents. So, if the age of some astronomical object (at some
redshift) is measured accurately, then it can be used to test
cosmological models according to this simple age principle. Now,
there are some OHROs discovered, for example, the $3.5$ Gyr old
galaxy LBDS 53W091 at redshift $z=1.55$~\cite{r46} and the $4.0$ Gyr
old galaxy LBDS 53W069 at redshift $z=1.43$~\cite{r48}. In
particular, the old quasar APM $08279+5255$ at redshift $z=3.91$ is
an important one, which has been used as a ``cosmic clock'' to
constrain cosmological models. Its age is estimated to be $2.0-3.0$
Gyr \cite{r51}. These three OHROs at $z=1.43$, 1.55 and 3.91 have
been used to test many dark energy models, including the
$\Lambda$CDM model~\cite{r44}, the general EoS dark energy
model~\cite{r57}, the scalar-tensor quintessence model~\cite{r58},
the $f(R)=\sqrt{R^2-R_0^2}$ model~\cite{r59}, the DGP braneworld
model~\cite{r60}, the power-law parameterized quintessence
model~\cite{r62}, the Yang-Mills condensate
model~\cite{Tong:2009mu}, the holographic dark energy
model~\cite{hao}, the agegraphic dark energy
model~\cite{Zhang:2007ps}, and so on. These investigations show that
the two OHROs at $z=1.43$ and 1.55 can be easily accommodated in
most dark energy models, whereas the OHRO at $z=3.91$ cannot, even
in the $\Lambda$CDM model~\cite{r44} and the holographic dark energy
model ~\cite{hao}.

In this Letter, we revisit the cosmic age problem in the holographic
dark energy model. We consider an interacting holographic dark
energy model in a non-flat universe. We will show that the age
crisis in the original holographic dark energy model can be avoided
when the interaction and the spatial curvature are involved in the
holographic dark energy model.


\section{The holographic dark energy model with spatial curvature and
 interaction}\label{sec2}

In this section we describe the interacting holographic dark energy
in a non-flat universe. In a spatially non-flat
Friedmann-Robertson-Walker (FRW) universe, the Friedmann equation
reads
 \be\label{eq3}
 \ 3M_{Pl}^2H^2=\rho_\Lambda+\rho_m-{3M_{Pl}^2k \over a^2},
\ee where $\rho_\Lambda=3c^2M_{Pl}^2L^{-2}$ is the holographic dark
energy density, and $\rho_m$ is the energy density of matter. We
define \be\label{eq4}
 \Omega_k=-{k \over H^2a^2}=\Omega_{k0}\Big({H_0\over
aH}\Big)^2,\ \ \ \ \ \Omega_{\Lambda}={\rho_{\Lambda} \over
\rho_c},\ \ \ \ \ \Omega_{m}={\rho_{m} \over \rho_c}, \ee
 where
$\rho_c=3M_{Pl}^2H^2$ is the critical density of the universe, thus
we have
 \be\label{eq5}
  \Omega_m+\Omega_{\Lambda}+\Omega_k=1.
 \ee

Now, let us consider some interaction between holographic dark
energy and matter:
 \be\label{eq6} \dot\rho_m+3H\rho_m=Q, \ee
 \be\label{eq7} \dot\rho_{\Lambda}+3H(\rho_{\Lambda}+p_{\Lambda})=-Q, \ee
where $Q$ denotes the phenomenological interaction term. Owing to
the lack of the knowledge of micro-origin of the interaction, we
simply follow other work on the interacting holographic dark energy
and parameterize the interaction term generally as
$Q=3H(\alpha\rho_{\Lambda}+\beta\rho_m)$, where $\alpha$ and $\beta$
are the dimensionless coupling constants. For reducing the
complication and the number of parameters, one often considers the
following three cases: (i) $\beta=0$, and thus $Q=3\alpha
H\rho_\Lambda$, (ii) $\alpha=\beta$, and thus $Q=3\alpha
H(\rho_\Lambda+\rho_m)$, and (iii) $\alpha=0$, and thus $Q=3\beta
H\rho_m$. Note that in these three cases, according to our
convention, $\alpha>0$ (or $\beta>0$) means that dark energy decays
to matter. Moreover, it should be pointed out that $\alpha<0$ (or
$\beta<0$) will lead to unphysical consequences in physics, since
$\rho_m$ will become negative and $\Omega_\Lambda$ will be greater
than 1 in the far future. So, in the present Letter, we only
consider the physically reasonable situations, namely, $\alpha>0$ or
$\beta>0$ in the above three cases. In the rest of this section, we
will formulate the model generally (by using both $\alpha$ and
$\beta$), but in the next section we will only consider the above
three simpler cases due to the aforementioned reason.

From the definition of holographic dark energy (\ref{eq1}), we have
\be\label{eq8} \Omega_{\Lambda}={c^2\over H^2L^2}, \ee or
equivalently,
\be\label{eq9}
 L={c \over H\sqrt{\Omega_{\Lambda}}}.
\ee Thus, we easily get
\be\label{eq10}
 \dot L=-{c
\over H \sqrt{\Omega_{\Lambda}}}\left({\dot H \over
H}+{\dot{\Omega}_{\Lambda}\over 2\Omega_{\Lambda}} \right).
 \ee

Following Ref.~\cite{Huang:2004ai}, in a non-flat universe the IR
cutoff length scale $L$ takes the form \be\label{eq11}
 L=ar(t),
\ee and $r(t)$ satisfies \be\label{eq12}
  \int_0^{r(t)} {dr \over
\sqrt{1-kr^2}}=\int_t^{+\infty}{dt\over a(t)}. \ee Consequently, we
have
 \be\label{eq13}
 r(t)={1\over\sqrt{k}}\sin
\Big(\sqrt{k}\int_t^{+\infty} {dt \over a}\Big)={1\over\sqrt{k}}\sin
\Big(\sqrt{k}\int_{a(t)}^{+\infty} {da \over {Ha^2}}\Big). \ee
Equation~(\ref{eq11}) leads to another equation about $r(t)$,
namely,
 \be\label{eq14}
 r(t)={L\over
a}={c\over\sqrt{\Omega_{\Lambda}}Ha}. \ee Combining
Eqs.~(\ref{eq13}) and (\ref{eq14}) yields \be\label{eq15}
\sqrt{k}\int_t^{+\infty}{dt\over a}=\arcsin{c\sqrt{k}\over
\sqrt{\Omega_{\Lambda}}aH}. \ee Taking the derivative of
Eq.~(\ref{eq15}) with respect to $t$, one can get \be\label{eq16}
\sqrt{{\Omega_{\Lambda}H^2\over c^2}-{k\over
a^2}}={\dot\Omega_{\Lambda}\over2\Omega_{\Lambda}}+H+{\dot H\over
H}. \ee

Let us combine Eqs.~(\ref{eq6}) and (\ref{eq7}), and then we have
${(\dot{\rho}_{\Lambda}+\dot{\rho}_m)}+3H(\rho_{\Lambda}+\rho_m+p_{\Lambda})=0$,
which is equivalent to the equation
${(\dot{\rho}_{c}-\dot{\rho}_k)}+3H(\rho_{c}-\rho_k+p_{\Lambda})=0$.
From this equation, we can obtain the form of $p_\Lambda$:
\begin{equation}
p_{\Lambda}=-{1\over3H}\Big(2{\dot H\over H}\rho_c+2{\dot a\over
a}\rho_k \Big)-\rho_c+\rho_k.\label{eq17}
\end{equation}
On the other hand, from Eqs.~(\ref{eq1}), (\ref{eq9}) and
(\ref{eq10}), we find that \be\label{eq18}
 \dot{\rho}_{\Lambda}=2\rho_{\Lambda}\left({\dot\Omega_{\Lambda}\over2\Omega_{\Lambda}}+{\dot H\over
H}\right). \ee Furthermore, substituting Eqs.~(\ref{eq17}) and
(\ref{eq18}) into Eq.~(\ref{eq7}), we obtain \be\label{eq19}
\dot{\Omega}_{\Lambda}+2{\dot{H}\over{H}}(\Omega_{\Lambda}-1)+H(3\Omega_{\Lambda}-3+\Omega_{k})
=-3H(\alpha\Omega_{\Lambda}+\beta\Omega_{m}).\ee
\begin{widetext}Combining this equation with Eq.~(\ref{eq16}), we
eventually obtain the following equations governing the dynamical
evolution of the interacting holographic dark energy in a non-flat
universe,
\begin{equation}\label{eq20}
{1\over H/H_0}{d\over dz}\left({H\over H_0}\right)
=-{\Omega_{\Lambda}\over
1+z}\left({\Omega_{\Lambda}-3+{\Omega_{k0}(1+z)^2\over(H/H_0)^2}
+3\alpha\Omega_{\Lambda}+3\beta(1-\Omega_{\Lambda}-{\Omega_{k0}(1+z)^2\over(H/H_0)^2})
\over2\Omega_{\Lambda}}+\sqrt{{\Omega_{\Lambda}\over c^2}
+{\Omega_{k0}(1+z)^2\over (H/H_0)^2}} \right),
\end{equation}
\be\label{eq21} {d\Omega_{\Lambda}\over
dz}=-{\Omega_{\Lambda}(1-\Omega_{\Lambda})\over
1+z}\left(2\sqrt{{\Omega_{\Lambda}\over
c^2}+{\Omega_{k0}(1+z)^2\over(H/H_0)^2}}+1-{3\alpha\Omega_{\Lambda}+{(1+z)^2\Omega_{k0}\over(H/
H_0)^2}+3\beta(1-\Omega_{\Lambda}-{\Omega_{k0}(1+z)^2\over(H/H_0)^2})\over
1-\Omega_{\Lambda}}\right). \ee These two equations can be solved
numerically, and the solutions, $\Omega_\Lambda (z)$ and $H(z)$,
determine the expansion history of the universe in the holographic
dark energy model.\end{widetext}

The holographic dark energy model with spatial curvature and
interaction described in this section has been strictly constrained
in Ref.~\cite{Li:2009zs} by using the current observational data
including the SNIa Constitution data, the shift parameter of the CMB
given by the five-year WMAP observations, and the BAO measurement
from the SDSS. The main fitting results were summarized as Table~I
and Figs.~1-5 of Ref.~\cite{Li:2009zs}. In the following
discussions, we restrict the values of parameters to the
observational constraint results derived by Ref.~\cite{Li:2009zs}.
Note that our definition of $\Omega_k$, $\alpha$ and $\beta$ are
different from that of Ref.~\cite{Li:2009zs} by a minus sign.


\section{Testing the model with the OHRO}\label{sec3}

The age of the universe at redshift $z$ is given by \be\label{eq22}
t(z)=\int_z^\infty\frac{dz'}{(1+z')H(z')}. \ee For convenience, we
introduce the dimensionless cosmic age
 \be\label{eq23}
 T_{cos}(z)\equiv H_0 t(z)=\int_z^\infty
 \frac{dz'}{(1+z')E(z')},
 \ee
where $E(z)\equiv H(z)/H_0$, and for the holographic dark energy
model it is given by the solutions of Eqs.~(\ref{eq20}) and
(\ref{eq21}). At any redshift, the age of the universe should be
larger than, or at least equal to, the age of the OHRO, namely
$T_{cos}(z)\geq T_{obj}(z)\equiv H_0 t_{obj}(z)$, where $t_{obj}(z)$
is the age of the OHRO at redshift $z$. Following Ref.~\cite{hao},
we define a dimensionless quantity, the ratio of the cosmic age and
the OHRO age,
\begin{equation}
\tau(z)\equiv {T_{cos}(z)\over
T_{obj}(z)}=H_0^{-1}t_{obj}^{-1}(z)\int_z^\infty
\frac{dz'}{(1+z')E(z')}.\label{tau}
\end{equation}
So, the condition $T_{cos}(z)\geq T_{obj}(z)$ is translated into
$\tau(z)\geq 1$. From Eq.~(\ref{tau}), it is easy to see that given
the age of OHRO $t_{obj}(z)$, the lower $H_0$, the higher $\tau(z)$;
given the Hubble constant $H_0$, the smaller $t_{obj}(z)$, the
larger $\tau(z)$.

In the work of Wei and Zhang~\cite{hao}, the original holographic
dark energy model (neither spatial curvature nor interaction is
involved) has been examined by using the three OHROs, the old galaxy
LBDS 53W091 at redshift $z=1.55$, the old galaxy LBDS 53W069 at
redshift $z=1.43$, and the old quasar APM $08279+5255$ at redshift
$z=3.91$. It is found in Ref.~\cite{hao} that the former two OHROs,
the old galaxy LBDS 53W091 at redshift $z=1.55$ and the old galaxy
LBDS 53W069 at redshift $z=1.43$, can be easily accommodated, but
the last one, the old quasar APM $08279+5255$ at redshift $z=3.91$,
cannot be accommodated in the model. In the present Letter, we
extend the holographic dark energy model to involving the spatial
curvature and the interaction, as described in the previous section,
and we shall examine whether the OHRO, the old quasar APM
$08279+5255$ at redshift $z=3.91$, is consistent with such a
sophisticated holographic dark energy model.

For the age of the OHRO at $z=3.91$, following Ref.~\cite{hao}, we
use the lower bound estimated, $t_{obj}(3.91)=2.0$ Gyr. For the
holographic dark energy model, since the main goal of this Letter is
to probe the effects of spatial curvature and interaction in
fighting against the cosmic age crisis, we keep the values of $c$
and $\Omega_{m0}$ fixed in the whole Letter. We take $c=0.8$ and
$\Omega_{m0}=0.28$ that are consistent with the observational
constraint results of Ref.~\cite{Li:2009zs}. For decreasing the
complication, let us close some parameters in turn. We shall
consider the following three cases: (a) the model of holographic
dark energy with spatial curvature but without interaction (namely,
$\Omega_{k0}\neq 0$ but $Q=0$), denoted as KHDE; (b) the model of
holographic dark energy with interaction but without spatial
curvature (namely, $Q\neq 0$ but $\Omega_{k0}=0$), denoted as IHDE;
(c) the model of holographic dark energy with both interaction and
spatial curvature (namely, $Q\neq 0$ and $\Omega_{k0}\neq 0$),
denoted as KIHDE. Next, let us discuss the use of the Hubble
constant $H_0=100 h$ km s$^{-1}$ Mpc$^{-1}$. Based on the HST key
project, Freedman et al.~\cite{r64} give the result $h=0.72\pm
0.08$. However, recently, many authors argue for a lower Hubble
constant, say, $h=0.68\pm 0.07$ (2$\sigma$)~\cite{r72}. Moreover,
the final result of the 15-year HST program given by Sandage et
al.~\cite{Sandage} is $h=0.623\pm 0.063$ which has attracted more
and more attention. Furthermore, when the holographic dark energy
model is fitted via observational data (SNIa+BAO+CMB), a lower value
of $h$ ($h\sim 0.65$) is obtained~\cite{Li:2009zs} (the latest fit
value is $h=0.686$ \cite{Li:2009jx}). It should also be mentioned
that the result of the 7-year WMAP observations (WMAP+BAO+$H_0$) is
$h=0.704^{+0.013}_{-0.014}$ \cite{Komatsu:2010fb}, which is derived
based on a $\Lambda$CDM model. In this Letter, we follow
Ref.~\cite{Li:2009zs} and take $h=0.64$ that is the lower bound of
Freedman et al.~\cite{r64}. We will also extend our discussion by
taking some higher values of $h$ into account (say, we will also
consider $h=0.72$, the central value of Freedman et al.~\cite{r64},
which is high enough for our discussion, since it is even higher
than the upper bound of WMAP 7-year result). Note that
$T_{obj}(3.91)=0.131$ is obtained according to $t_{obj}(3.91)=2.0$
Gyr and $h=0.64$.

\begin{center}
 \begin{figure}[htbp]
  \centering
 \includegraphics[width=0.4\textwidth]{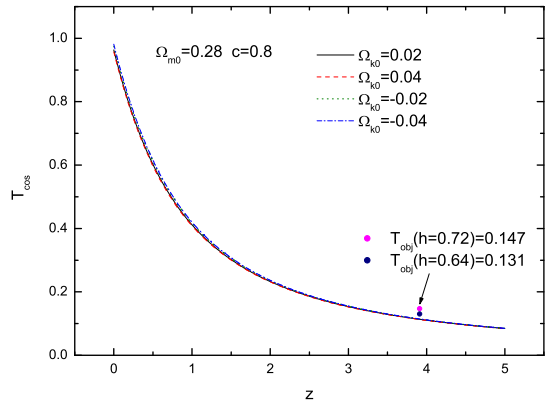}
\caption{\label{fig:KHDE} The dimensionless cosmic age $T_{cos}(z)$
in the KHDE model. For the curves, we fix $c=0.8$,
$\Omega_{m0}=0.28$ and $h=0.64$, and take $\Omega_{k0}=0.02$, 0.04,
$-0.02$ and $-0.04$. The dots represent the dimensionless age of the
old quasar APM $08279+5255$ at $z=3.91$, $T_{obj}$, under the
assumption $t_{obj}=2.0$ Gyr. The blue dot corresponds to $h=0.64$
and the pink one corresponds to $h=0.72$.}

 \end{figure}
 \end{center}

First, we test the KHDE model. The current observational constraint
result of the KHDE model is~\cite{Li:2009zs}: $-0.02\lesssim
\Omega_{k0}\lesssim 0.02$ (1$\sigma$). When we take
$\Omega_{k0}=0.02$, we find $\tau(3.91)=0.866$, less than 1; when we
take $\Omega_{k0}=-0.02$, we obtain $\tau(3.91)=0.872$, still less
than 1. So, we find that the spatial curvature is hard to help solve
the cosmic age crisis for the holographic dark energy model. From
the above example, we find that the value of $\tau$ in a closed
space is greater than that in an open space. Thus, let us increase
the value of $|\Omega_{k0}|$ in a closed space geometry in order to
see whether the problem can be solved in some extremal cases. Our
efforts can be found in Table~\ref{tab:khde}. In this table, we see
that even the value of $\Omega_{k0}$ is taken to be $-0.1$, the
value of $\tau$ derived is merely 0.883, far from solving the cosmic
age problem. In addition, we also plot the $T_{cos}(z)$ curves for
the KHDE model in Fig.~\ref{fig:KHDE}. It can be explicitly seen
from this figure that the cosmic age problem is still acute in the
KHDE model. Therefore, the conclusion is that the cosmic age crisis
cannot be avoided by only considering the spacial curvature in the
holographic dark energy model.

 \begin{table}[htbp]
 \begin{center}
 \caption{\label{tab1} The ratio $\tau (3.91)\equiv T_{cos}(3.91)/T_{obj}(3.91)$
 for different $\Omega_{k0}$ in the KHDE model with $c=0.8$ and $\Omega_{m0}=0.28$.}\label{tab:khde}
 \begin{tabular}{ccccccc} \toprule[0.8pt]
 $\Omega_{k0}$   &   $0.04$   &   $0.02$   &   $-0.02$   &   $-0.04$   &   $-0.06$   &   $-0.1$\\ \hline
 $T_{cos}(3.91)$   &   $0.1131$   &   $0.1135$   &   $0.1143$   &   $0.1146$   &   $0.1150$   &   $0.1157$  \\\hline
 $\tau(3.91)$   &   $0.864$   &   $0.866$   &   $0.872$   &   $0.875$   &   $0.878$   &   $0.883$  \\
 \bottomrule[0.8pt]
 \end{tabular}
 \end{center}
 \end{table}

\begin{center}
 \begin{figure}[htbp]
 \centering
 \includegraphics[width=0.4\textwidth]{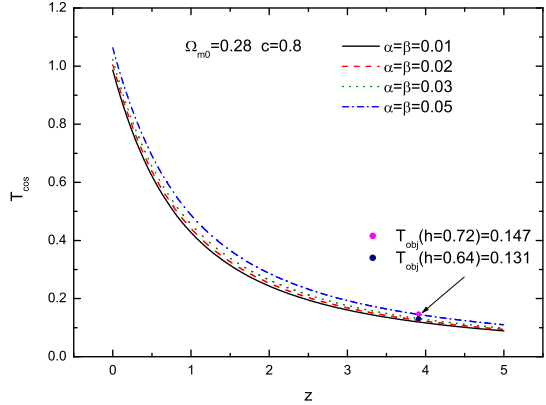}
 \caption{\label{fig:IHDE} The dimensionless cosmic age $T_{cos}(z)$
in the IHDE2 model. For the curves, we fix $c=0.8$,
$\Omega_{m0}=0.28$ and $h=0.64$, and take $\alpha=\beta=0.01$, 0.02,
0.03 and 0.05. The dots represent the dimensionless age of the old
quasar APM $08279+5255$ at $z=3.91$, $T_{obj}$, under the assumption
$t_{obj}=2.0$ Gyr. The blue dot corresponds to $h=0.64$ and the pink
one corresponds to $h=0.72$.}
 \end{figure}
 \end{center}

For the IHDE model, we consider the aforementioned three cases: (I)
$\beta=0$, named IHDE$1$; (II) $\alpha=\beta$, named IHDE$2$; (III)
$\alpha=0$, named IHDE$3$. To see how the interaction influences the
cosmic age in the holographic dark energy model, we calculate the
age for these three cases in Table~\ref{tab2}. From this table, we
see that with the increase of the interaction parameter $\alpha$ or
$\beta$, the cosmic age $T_{cos}$ also increases. It is clear that
the value of $\tau(3.91)$ can be greater than 1 when the value of
$\alpha$ (or $\beta$) is large enough. For example, for the case of
IHDE2 (Case II), when $\alpha$ is taken to be 0.03, the value of
$\tau(3.91)$ obtained is 1.005. The cosmic age $T_{cos}$ versus
redshift $z$ in the case of IHDE2 is also displayed in
Fig.~\ref{fig:IHDE}. This figure shows explicitly that the age
problem can be overcomed when the interaction is involved in the
holographic dark energy model. However, it should be pointed out
that the parameter values making $\tau(3.91)>1$ actually exceed the
$2\sigma$ regions given by Ref.~\cite{Li:2009zs}. Therefore, if we
confine our discussions in the parameter space constrained by
current observational data, the problem is not so easy as it looks.
Nevertheless, it is found in Ref.~\cite{Li:2009zs} that when
simultaneously considering the interaction and spatial curvature in
the holographic dark energy model, the parameter space is amplified,
especially, the ranges of $\alpha$ (or $\beta$) and $\Omega_{k0}$
are enlarged by 10 times comparing to the IHDE and KHDE models.
Based on this fact, it can be expected that the age problem could be
solved when the interaction and spatial curvature are both taken
into account.

\begin{table}[htbp]
\begin{center}
\caption{\label{tab2} The values of $T_{cos}(3.91)$ and $\tau(3.91)$
in the IHDE models with $c=0.8$, $\Omega_{m0}=0.28$ and $h=0.64$.}
\begin{tabular}{cccccc}\toprule[0.8pt]
 Case~I  ~$(\beta=0)$&  ~$\alpha$  & ~ $0.02$  &  ~$0.06$  &  ~$0.10$  &  ~$0.15$\\
 \cline{2-6}
  &$T_{cos}(3.91)$   &   $0.1172$   &   $0.1246$   &   $0.1335$   &   $0.1475$ \\\cline{2-6}
   &  ~$\tau(3.91)$  &  ~$0.894$  &  ~$0.951$  &  ~$1.019$  &  ~$1.126$  \\ \bottomrule[0.8pt]
Case~II~$(\alpha=\beta)$  &  ~$\alpha$  &  ~$0.01$  &  ~$0.02$  &  ~$0.03$  &  ~$0.05$\\
\cline{2-6}
 &$T_{cos}(3.91)$   &   $0.1194$   &   $0.1253$   &   $0.1316$   &   $0.1456$ \\\cline{2-6}
   &  ~$\tau(3.91)$  &  ~$0.912$  &  ~$0.957$  &  ~$1.005$  &  ~$1.111$  \\ \bottomrule[0.8pt]
 Case~III~$(\alpha=0)$  &  ~$\beta~$  &  ~$0.01$  &  ~$0.03$  &  ~$0.05$  &  ~$0.07$\\ \cline{2-6}
  &$T_{cos}(3.91)$   &   $0.1177$   &   $0.1259$   &   $0.1346$   &   $0.1440$ \\\cline{2-6}
  &  ~$\tau(3.91)$  &  ~$0.899$  &  ~$0.961$  &  ~$1.028$  &  ~$1.099$  \\ \bottomrule[0.8pt]
\end{tabular}
\end{center}
\end{table}

\begin{center}
 \begin{figure}[htbp]
 \centering
 \includegraphics[width=0.4\textwidth]{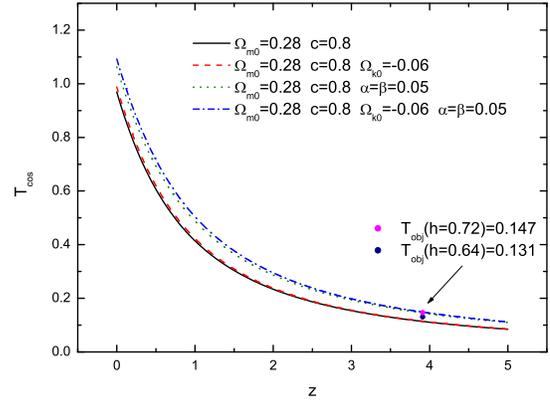}
 \caption{\label{fig:KIHDE} The dimensionless cosmic age $T_{cos}(z)$
in the KIHDE2 model. For the curves, we fix $c=0.8$,
$\Omega_{m0}=0.28$ and $h=0.64$. The sophisticated case of
interacting holographic dark energy in a non-flat universe is
represented by the blue dash-dotted curve (where we take
$\Omega_{k0}=-0.06$ and $\alpha=\beta=0.05$), and other cases such
as HDE (black curve), KHDE (red dashed curve) and IHDE (green dotted
curve) are the special cases in this framework. The dots represent
the dimensionless age of the old quasar APM $08279+5255$ at
$z=3.91$, $T_{obj}$, under the assumption $t_{obj}=2.0$ Gyr. The
blue dot corresponds to $h=0.64$ and the pink one corresponds to
$h=0.72$..}
 \end{figure}
 \end{center}

Now, let us consider the KIHDE model. For simplicity, in our
discussion we fix $\Omega_{k0}=-0.06$. The three phenomenological
interaction cases are the same as in the IHDE model. Since the
parameter space of the KIHDE model is greatly amplified, the
interaction parameter can be chosen to be some large values. For
example, for the KIHDE2 case, one can choose $\alpha=0.05$ that is
allowed by current observations, then the result $\tau(3.91)=1.137$
is obtained. Some typical examples for all the three cases are shown
in Table~\ref{tab3}, where the values of the interaction parameters
are taken within the $2\sigma$ ranges of the observational
constrains given by Ref.~\cite{Li:2009zs}. It is explicitly shown
that the cosmic age problem can be successfully solved in the KIHDE
model. For clarity, we plot the curves of $T_{cos}(z)$ in
Fig.~\ref{fig:KIHDE}. This figure shows a direct comparison of HDE,
KHDE, IHDE, and KIHDE (the Case II of interaction is taken as an
example in this figure). It should be noted that $\alpha=0.05$ is
not allowed in the IHDE model but is allowed in the KIHDE model,
from the viewpoint of observation. From Fig.~\ref{fig:KIHDE}, we
also see that the age problem can be evaded in the KIHDE even a much
larger value of $h$ is taken, for instance, when $h=0.72$, we get
$T_{obj}(3.91)=0.147$, $T_{cos}(3.91)=0.149$, and thus $\tau>1$ in
this case. Moreover, when a larger age of the quasar is taken, say,
$t_{obj}(3.91)=2.1$ Gyr, the age problem can also be overcame in the
KIHDE model; in Table~\ref{tab4} one can find the values of
$T_{obj}(3.91)$ corresponding to $t_{obj}(3.91)=2.1$ Gyr for
$h=0.64$ and 0.72. Therefore, the cosmic age crisis can be avoided
in the holographic dark energy model when the interaction and
spatial curvature are both taken into account. Nevertheless, we have
to admit that the price for solving the age problem has been paid by
the holographic dark energy model, i.e., there are too many free
parameters have to be considered in the model. This would inevitably
weaken, to some extent, the plausibility of the model.

\begin{table}[htbp]
\begin{center}
\caption{\label{tab3} The values of $T_{cos}(3.91)$ and $\tau(3.91)$
in the KIHDE models with $c=0.8$, $\Omega_{m0}=0.28$,
$\Omega_{k0}=-0.06$ and $h=0.64$.}
\begin{tabular}{c c c c} \toprule[0.8pt]
 Case~I~$(\beta=0)$ &  ~$\alpha$  &  ~$0.1$  &  ~$0.2$\\\cline{2-4}
  &$T_{cos}(3.91)$   &   $0.1375$   &   $0.1782$\\\cline{2-4}
   &  ~$\tau(3.91)$  &  ~$1.049$  &  ~$1.360$\\ \bottomrule[0.8pt]
 Case~II~$(\alpha=\beta)$  &  ~$\alpha$  &  ~$0.05$  &  ~$0.1$\\ \cline{2-4}
 &$T_{cos}(3.91)$   &   $0.1489$   &   $0.1993$\\\cline{2-4}
   &  ~$\tau(3.91)$  &  ~$1.137$  &  ~$1.521$\\ \bottomrule[0.8pt]
 Case~III~$(\alpha=0)$ &  ~$\beta$  &  ~$0.05$  &  ~$0.1$\\ \cline{2-4}
 &$T_{cos}(3.91)$   &   $0.1364$   &   $0.1623$\\\cline{2-4}
   &  ~$\tau(3.91)$  & ~ $1.041$  &  ~$1.239$\\ \bottomrule[0.8pt]
 \end{tabular}
 \end{center}
 \end{table}

Of course, to be honest, it should also be confessed that the age
problem would still exist if one considers some extremal cases such
as a much larger possible age of the quasar $t_{obj}$ with a larger
$h$. Consider the upper limit of the quasar age, $t_{obj}(3.91)=3.0$
Gyr. For this extreme case, when $h=0.64$, we have
$T_{obj}(3.91)=0.196$; when $h=0.72$, we have $T_{obj}(3.91)=0.220$;
see also Table~\ref{tab4}. So, we have to admit that for the limit
case of $t_{obj}(3.91)=3.0$ Gyr and $h=0.72$ the age problem cannot
be solved yet even in the KIHDE model.

 \begin{table}[htbp]
 \begin{center}
  \caption{\label{tab4} The values of $T_{obj}(3.91)$ corresponding to different $h$ and $t_{obj}(3.91)$.}
 \begin{tabular}{c c c} \toprule[0.8pt]
 $h$ &  ~$t_{obj}(3.91)$/Gyr  &  ~$T_{obj}(3.91)$\\\bottomrule[0.8pt]
  ~   &   $2.0$   &   $0.131$\\
0.64    &  ~$2.1$  &  ~$0.137$\\  
~  & ~$3.0$ &~$0.196$\\\bottomrule[0.8pt]
 ~  &   $2.0$   &   $0.147$\\
   0.72  &  ~$2.1$  &  ~$0.154$\\ 
~ &  ~$3.0$  &  ~$0.220$\\\bottomrule[0.8pt]

 \end{tabular}
 \end{center}
 \end{table}


\section{Conclusion}\label{sec4}

In this Letter, we have revisited the cosmic age problem in the
holographic dark energy model. The cosmic age problem brought by the
old quasar APM $08279+5255$ has caused trouble to many cosmological
models, and the holographic dark energy model is not an exception
either \cite{hao}. In order to accommodate the old quasar APM
$08279+5255$ in the holographic dark energy model, we propose to
consider the interaction between dark energy and matter in the
model. We have shown that the quasar indeed can be accommodated in
the holographic dark energy model when an appropriate interaction
strength is chosen. Taking the current observational constraints
\cite{Li:2009zs} into account, we have demonstrated that both
interaction and spatial curvature should be simultaneously involved
in the holographic dark energy model. It has been shown that if such
a sophisticated case is considered the quasar APM $08279+5255$ can
be accommodated and the cosmic age problem can thus be avoided in
the holographic dark energy model. The price of solving the age
problem in this way is also apparent, i.e., the model involves too
many free parameters, which may weaken the plausibility of the
model, to some extent.

It is well known that the consideration of interaction in the
holographic dark energy can be used to avoid the future big-rip
singularity caused by $c<1$ \cite{Li:2009zs,Li:2008zq}. In this
Letter we have provided another advantage for the consideration of
interaction in the holographic dark energy, i.e., the interaction
between dark energy and matter can also be used to avoid the age
problem caused by the old quasar. So, our result can be viewed as a
further support to the interacting holographic dark energy model.

Of course, we have to confess that the age problem would still exist
if some extreme cases are taken into account, say, a much larger
possible age of the quasar $t_{obj}$ with a larger $h$. It is
remarkable that the age of the old quasar APM $08279+5255$ has not
been measured accurately yet, and the age problem caused by this
quasar has troubled many dark energy models (including the
$\Lambda$CDM model). It is expected that the future accurate
measurement on the age of this old quasar would eliminate the cosmic
age crisis in dark energy models.


\begin{acknowledgments}
This work was supported by the Natural Science Foundation of China
under Grant Nos.~10705041 and 10975032.
\end{acknowledgments}



\begin{thebibliography}{99}
\bibitem{Riess98}
  A.~G.~Riess {\it et al.}  [Supernova Search Team Collaboration],
  Astron.\ J.\  {\bf 116}, 1009 (1998)
  [astro-ph/9805201];
  S.~Perlmutter {\it et al.}  [Supernova Cosmology Project Collaboration],
  Astrophys.\ J.\  {\bf 517}, 565 (1999)
  [astro-ph/9812133].

\bibitem{Tegmark04}
  M.~Tegmark {\it et al.}  [SDSS Collaboration],
  Phys.\ Rev.\ D {\bf 69}, 103501 (2004)
  [astro-ph/0310723];
  K.~Abazajian {\it et al.}  [SDSS Collaboration],
  Astron.\ J.\  {\bf 128}, 502 (2004)
  [astro-ph/0403325];
  K.~Abazajian {\it et al.}  [SDSS Collaboration],
  Astron.\ J.\  {\bf 129}, 1755 (2005)
  [astro-ph/0410239].

\bibitem{Spergel03}
D.~N.~Spergel {\it et al.}  [WMAP Collaboration],
  Astrophys.\ J.\ Suppl.\  {\bf 148} 175 (2003)
  [astro-ph/0302209].

\bibitem{dereview}
S.~Weinberg,
  Rev.\ Mod.\ Phys.\  {\bf 61} (1989) 1;
V.~Sahni and A.~A.~Starobinsky,
  Int.\ J.\ Mod.\ Phys.\  D {\bf 9} (2000) 373
  [arXiv:astro-ph/9904398];
S.~M.~Carroll,
  Living Rev.\ Rel.\  {\bf 4} (2001) 1
  [arXiv:astro-ph/0004075];
P.~J.~E.~Peebles and B.~Ratra,
  Rev.\ Mod.\ Phys.\  {\bf 75} (2003) 559
  [arXiv:astro-ph/0207347];
T.~Padmanabhan,
  Phys.\ Rept.\  {\bf 380} (2003) 235
  [arXiv:hep-th/0212290];
E.~J.~Copeland, M.~Sami and S.~Tsujikawa,
  Int.\ J.\ Mod.\ Phys.\  D {\bf 15} (2006) 1753
  [arXiv:hep-th/0603057].


\bibitem{quintessence}
  P.~J.~E.~Peebles and B.~Ratra,
  Astrophys.\ J.\  {\bf 325} L17 (1988);
  B.~Ratra and P.~J.~E.~Peebles,
  Phys.\ Rev.\ D {\bf 37} 3406 (1988);
  C.~Wetterich,
  Nucl.\ Phys.\ B {\bf 302} 668 (1988);
  I.~Zlatev, L.~M.~Wang and P.~J.~Steinhardt,
  Phys.\ Rev.\ Lett.\  {\bf 82}, 896 (1999)
  [astro-ph/9807002];
  X.~Zhang,
  Phys.\ Lett.\  B {\bf 648}, 1 (2007)
  [arXiv:astro-ph/0604484].



\bibitem{kessence}
  C.~Armendariz-Picon, V.~F.~Mukhanov and P.~J.~Steinhardt,
  Phys.\ Rev.\ Lett.\  {\bf 85}, 4438 (2000)
  [astro-ph/0004134];
  C.~Armendariz-Picon, V.~F.~Mukhanov and P.~J.~Steinhardt,
  Phys.\ Rev.\ D {\bf 63}, 103510 (2001)
  [astro-ph/0006373].


\bibitem{hessence}
H.~Wei, R.~G.~Cai and D.~F.~Zeng,
 Class.\ Quant.\ Grav.\  {\bf 22}, 3189 (2005) [hep-th/0501160];
H.~Wei and R.~G.~Cai,
 Phys.\ Rev.\ D {\bf 72}, 123507 (2005) [astro-ph/0509328];
M.~Alimohammadi and H.~Mohseni Sadjadi,
 Phys.\ Rev.\ D {\bf 73}, 083527 (2006) [hep-th/0602268];
W.~Zhao and Y.~Zhang,
 Phys.\ Rev.\ D {\bf 73}, 123509 (2006) [astro-ph/0604460].


\bibitem{phantom}
  R.~R.~Caldwell, M.~Kamionkowski and N.~N.~Weinberg,
  Phys.\ Rev.\ Lett.\  {\bf 91}, 071301 (2003)
  [astro-ph/0302506];
  Z.~K.~Guo, Y.~S.~Piao and Y.~Z.~Zhang,
  Phys.\ Lett.\  B {\bf 594}, 247 (2004)
  [arXiv:astro-ph/0404225].


\bibitem{tachyon}
  A.~Sen,
  JHEP {\bf 0207}, 065 (2002)
  [hep-th/0203265];
  T.~Padmanabhan,
  Phys.\ Rev.\ D {\bf 66}, 021301 (2002)
  [hep-th/0204150];
  J.~Zhang, X.~Zhang and H.~Liu,
  Phys.\ Lett.\  B {\bf 651}, 84 (2007)
  [arXiv:0706.1185 [astro-ph]].


\bibitem{quintom}
  B.~Feng, X.~L.~Wang and X.~M.~Zhang,
  Phys.\ Lett.\ B {\bf 607}, 35 (2005)
  [astro-ph/0404224];
  Z.~K.~Guo, Y.~S.~Piao, X.~M.~Zhang and Y.~Z.~Zhang,
  Phys.\ Lett.\ B {\bf 608}, 177 (2005)
  [astro-ph/0410654];
  X.~Zhang,
  Commun.\ Theor.\ Phys.\  {\bf 44}, 762 (2005).

\bibitem{ghost}
  N.~Arkani-Hamed, H.~C.~Cheng, M.~A.~Luty and S.~Mukohyama,
  JHEP {\bf 0405}, 074 (2004)
  [hep-th/0312099];
  F.~Piazza and S.~Tsujikawa,
  JCAP {\bf 0407}, 004 (2004)
  [hep-th/0405054];
  X.~Zhang,
  Phys.\ Rev.\  D {\bf 74}, 103505 (2006)
  [astro-ph/0609699];
  J.~Zhang, X.~Zhang and H.~Liu,
  Mod.\ Phys.\ Lett.\  A {\bf 23}, 139 (2008)
  [astro-ph/0612642];
  X.~Zhang,
  Phys.\ Rev.\  D {\bf 79}, 103509 (2009)
  [arXiv:0901.2262 [astro-ph.CO]].



\bibitem{intde}
  L.~Amendola,
  Phys.\ Rev.\  D {\bf 62}, 043511 (2000)
  [arXiv:astro-ph/9908023];
  D.~Comelli, M.~Pietroni and A.~Riotto,
  Phys.\ Lett.\  B {\bf 571}, 115 (2003)
  [arXiv:hep-ph/0302080];
  X.~Zhang,
  Mod.\ Phys.\ Lett.\  A {\bf 20}, 2575 (2005)
  [arXiv:astro-ph/0503072];
  X.~Zhang,
  Phys.\ Lett.\  B {\bf 611}, 1 (2005)
  [arXiv:astro-ph/0503075];
  R.~G.~Cai and A.~Wang,
  JCAP {\bf 0503}, 002 (2005)
  [arXiv:hep-th/0411025];
  J.~H.~He, B.~Wang and P.~Zhang,
  Phys.\ Rev.\  D {\bf 80}, 063530 (2009)
  [arXiv:0906.0677 [gr-qc]].


\bibitem{brane}
  C.~Deffayet, G.~R.~Dvali and G.~Gabadadze,
  Phys.\ Rev.\ D {\bf 65}, 044023 (2002)
  [astro-ph/0105068];
  V.~Sahni and Y.~Shtanov,
  JCAP {\bf 0311}, 014 (2003)
  [astro-ph/0202346];
  W.~L.~Guo and X.~Zhang,
  Phys.\ Rev.\  D {\bf 79}, 115023 (2009)
  [arXiv:0904.2451 [hep-ph]];
  X.~Zhang,
  Phys.\ Lett.\  B {\bf 683}, 81 (2010)
  [arXiv:0909.4940 [gr-qc]].

\bibitem{cg}
  A.~Y.~Kamenshchik, U.~Moschella and V.~Pasquier,
  Phys.\ Lett.\ B {\bf 511}, 265 (2001)
  [gr-qc/0103004];
  M.~C.~Bento, O.~Bertolami and A.~A.~Sen,
  Phys.\ Rev.\  D {\bf 66}, 043507 (2002)
  [gr-qc/0202064];
  X.~Zhang, F.~Q.~Wu and J.~Zhang,
  JCAP {\bf 0601}, 003 (2006)
  [astro-ph/0411221].

\bibitem{YMC}
  W.~Zhao and Y.~Zhang,
  Class.\ Quant.\ Grav.\  {\bf 23}, 3405 (2006)
  [arXiv:astro-ph/0510356];
  S.~Wang, Y.~Zhang and T.~Y.~Xia,
  JCAP {\bf 0810}, 037 (2008)
  [arXiv:0803.2760 [gr-qc]].




\bibitem{hp}
G.~'t~Hooft, [gr-qc/9310026]; L.~Susskind,
 J.\ Math.\ Phys.\  {\bf 36}, 6377 (1995) [hep-th/9409089];
R.~Bousso,
 Rev.\ Mod.\ Phys.\  {\bf 74}, 825 (2002) [hep-th/0203101].

\bibitem{r16}
A.~G.~Cohen, D.~B.~Kaplan and A.~E.~Nelson,
 Phys.\ Rev.\ Lett.\  {\bf 82}, 4971 (1999)
 [hep-th/9803132].

 \bibitem{r17}
S.~D.~H.~Hsu,
 Phys.\ Lett.\  B {\bf 594}, 13 (2004) [hep-th/0403052].

\bibitem{Li04}
M.~Li,
  Phys.\ Lett.\  B {\bf 603} (2004) 1
  [hep-th/0403127].

\bibitem{Li:2009pm}
  M.~Li, R.~X.~Miao and Y.~Pang,
  Phys.\ Lett.\  B {\bf 689}, 55 (2010)
  [arXiv:0910.3375 [hep-th]].

\bibitem{r38}
Q.~G.~Huang and Y.~G.~Gong,
 JCAP {\bf 0408}, 006 (2004) [astro-ph/0403590].

\bibitem{holodata}
  X.~Zhang and F.~Q.~Wu,
  Phys.\ Rev.\  D {\bf 72}, 043524 (2005)
  [arXiv:astro-ph/0506310];
  X.~Zhang and F.~Q.~Wu,
  Phys.\ Rev.\  D {\bf 76}, 023502 (2007)
  [arXiv:astro-ph/0701405];
  M.~Li, X.~D.~Li, S.~Wang and X.~Zhang,
  JCAP {\bf 0906}, 036 (2009)
  [arXiv:0904.0928 [astro-ph.CO]].

\bibitem{Li:2009zs}
  M.~Li, X.~D.~Li, S.~Wang, Y.~Wang and X.~Zhang,
  JCAP {\bf 0912}, 014 (2009)
  [arXiv:0910.3855 [astro-ph.CO]].




\bibitem{r40}
Z.~Chang, F.~Q.~Wu and X.~Zhang,
 Phys.\ Lett.\  B {\bf 633}, 14 (2006) [astro-ph/0509531].

\bibitem{r41}
Z.~L.~Yi and T.~J.~Zhang,
 Mod.\ Phys.\ Lett.\  A {\bf 22}, 41 (2007) [astro-ph/0605596].

\bibitem{r42}
H.~Zhang, W.~Zhong, Z.~H.~Zhu and S.~He,
  Phys.\ Rev.\  D {\bf 76}, 123508 (2007)
  [arXiv:0705.4409 [astro-ph]].


\bibitem{Shen:2004ck}
See also, {\it e.g.},
  J.~Y.~Shen, B.~Wang, E.~Abdalla and R.~K.~Su,
  Phys.\ Lett.\  B {\bf 609}, 200 (2005)
  [arXiv:hep-th/0412227];
  Y.~Z.~Ma, Y.~Gong and X. Chen,
  Eur.\ Phys.\ J.\  C {\bf 60}, 303 (2009)
  [arXiv:0711.1641 [astro-ph]];
  Q.~Wu, Y.~Gong, A.~Wang and J.~S.~Alcaniz,
  Phys.\ Lett.\  B {\bf 659}, 34 (2008)
  [arXiv:0705.1006 [astro-ph]].




\bibitem{hao} H. Wei and S. N. Zhang, Phys. Rev. D \textbf{76}, 063003 (2007) [astro-ph/0707.2129].


\bibitem{r46}
J.~Dunlop {\it et al.}, Nature {\bf 381}, 581 (1996);
H.~Spinrad {\it et al.},
 Astrophys.\ J.\  {\bf 484}, 581 (1997).

\bibitem{r48}
J.~Dunlop, in {\it The Most Distant Radio Galaxies},
 edited by H.~J.~A.~Rottgering, P.~Best and M.~D.~Lehnert,
 Kluwer, Dordrecht  p.~71 (1999).

\bibitem{r51}
G.~Hasinger, N.~Schartel and S.~Komossa,
 Astrophys.\ J.\  {\bf 573}, L77 (2002) [astro-ph/0207005];
S.~Komossa and G.~Hasinger, [astro-ph/0207321].

\bibitem{r44}
J.~S.~Alcaniz and J.~A.~S.~Lima,
 Astrophys.\ J.\  {\bf 521}, L87 (1999) [astro-ph/9902298];
A.~Friaca, J.~Alcaniz and J.~A.~S.~Lima,
 Mon.\ Not.\ Roy.\ Astron.\ Soc.\  {\bf 362}, 1295 (2005)
 [astro-ph/0504031];
J.~S.~Alcaniz, J.~A.~S.~Lima and J.~V.~Cunha,
 Mon.\ Not.\ Roy.\ Astron.\ Soc.\  {\bf 340}, L39 (2003)
 [astro-ph/0301226].

\bibitem{r57}
M.~A.~Dantas, J.~S.~Alcaniz, D.~Jain and A.~Dev,
 Astron.\ Astrophys.\  {\bf 467}, 421 (2007)
 [astro-ph/0607060];
  D.~Jain and A.~Dev,
  Phys.\ Lett.\  B {\bf 633}, 436 (2006)
  [arXiv:astro-ph/0509212].



\bibitem{r58}
S.~Capozziello, P.~K.~S.~Dunsby, E.~Piedipalumbo and C.~Rubano,
  [astro-ph/0706.2615].

\bibitem{r59}
M.~S.~Movahed, S.~Baghram and S.~Rahvar, [astro-ph/0705.0889].

\bibitem{r60}
M.~S.~Movahed, M.~Farhang and S.~Rahvar, [astro-ph/0701339];
M.~S.~Movahed and S.~Ghassemi, [astro-ph/0705.3894];
N.~Pires, Z.~H.~Zhu and J.~S.~Alcaniz,
 Phys.\ Rev.\  D {\bf 73}, 123530 (2006) [astro-ph/0606689].

\bibitem{r62}
S.~Rahvar and M.~S.~Movahed,
 Phys.\ Rev.\  D {\bf 75}, 023512 (2007) [astro-ph/0604206].

\bibitem{Tong:2009mu}
S.~Wang and Y.~Zhang,
  Phys.\ Lett.\  B {\bf 669}, 201 (2008)
  [arXiv:0809.3627 [astro-ph]];
  M.~L.~Tong and Y.~Zhang,
  Phys.\ Rev.\  D {\bf 80}, 023503 (2009)
  [arXiv:0906.3646 [gr-qc]].

\bibitem{Zhang:2007ps}
  Y.~Zhang, H.~Li, X.~Wu, H.~Wei and R.~G.~Cai,
  arXiv:0708.1214 [astro-ph].

\bibitem{Huang:2004ai}
  Q.~G.~Huang and M.~Li,
  JCAP {\bf 0408}, 013 (2004)
  [arXiv:astro-ph/0404229].



\bibitem{r64}
W.~L.~Freedman {\it et al.},
 Astrophys.\ J.\  {\bf 553}, 47 (2001) [astro-ph/0012376].

\bibitem{r72}
J.~R.~I.~Gott, M.~S.~Vogeley, S.~Podariu and B.~Ratra,
 Astrophys.\ J.\  {\bf 549}, 1 (2001) [astro-ph/0006103];
G.~Chen, J.~R.~I.~Gott and B.~Ratra,
 Publ.\ Astron.\ Soc.\ Pac.\  {\bf 115}, 1269 (2003) [astro-ph/0308099].

\bibitem{Sandage}
A.~Sandage, G.~A.~Tammann, A.~Saha, B.~Reindl,
 F.~D.~Macchetto and N.~Panagia,
 Astrophys.\ J.\  {\bf 653}, 843 (2006) [astro-ph/0603647].

\bibitem{Li:2009jx}
  M.~Li, X.~D.~Li and X.~Zhang,
  arXiv:0912.3988 [astro-ph.CO].

\bibitem{Komatsu:2010fb}
  E.~Komatsu {\it et al.},
  arXiv:1001.4538 [astro-ph.CO].


\bibitem{Li:2008zq}
  M.~Li, C.~Lin and Y.~Wang,
  JCAP {\bf 0805}, 023 (2008)
  [arXiv:0801.1407 [astro-ph]].



\end{thebibliography}
\end{document}